\documentclass[12pt,a4paper]{article}
\usepackage[a4paper, inner=2.5cm, top=2.5cm, outer=2.5cm, bottom=2.5cm]{geometry}

\usepackage{graphicx}
\usepackage{bm}
\usepackage{subcaption} 
\usepackage{hyperref}
\usepackage{color}
\definecolor{darkblue}{rgb}{0,0,0.8}
\hypersetup{
	colorlinks=true,
	linkcolor=darkblue,
	citecolor=darkblue,
	urlcolor=darkblue
}
\usepackage{amssymb,amsfonts,amsmath,mathtext,cite,enumerate,float} 
\usepackage{color}
\usepackage{enumitem}
\usepackage{calc}
\usepackage{wrapfig}
\usepackage{authblk}

\usepackage{color}
\definecolor{alertcol}{rgb}{0.7,0, 0}
\definecolor{ascol}{rgb}{0.7,0, 0}
\definecolor{epcol}{rgb}{0,0.4, 0}
\definecolor{kgcol}{rgb}{0,0, 0.7}
\definecolor{incol}{rgb}{0.1,0.7, 0.7}

\newcommand{\angstrom}{\text{\normalfont\AA}}
\newcommand{\cfg}{{\rm cfg}}
\newcommand{\step}[1]{{\bf Step #1.} }
\begin{document}

\title{The MLIP package: Moment Tensor Potentials with MPI and Active Learning}
\author[1]{Ivan S. Novikov}
\author[1,2]{Konstantin Gubaev}
\author[1]{Evgeny V. Podryabinkin}
\author[1]{Alexander V. Shapeev}
\affil[1]{\normalsize Skolkovo Institute of Science and Technology, Skolkovo Innovation Center, Nobel St. 3, Moscow 143026, Russia}
\affil[2]{\normalsize Department of Materials Science and Engineering, Delft University of Technology, Mekelweg 2, 2628 CD Delft, Netherlands}
\maketitle

\subsection*{Abstract}
The subject of this paper is the technology (the ``how'') of constructing machine-learning interatomic potentials, rather than science (the ``what'' and ``why'') of atomistic simulations using machine-learning potentials.
Namely, we illustrate how to construct moment tensor potentials using active learning as implemented in the MLIP package, focusing on the efficient ways to sample configurations for the training set, how expanding the training set changes the error of predictions, how to set up ab initio calculations in a cost-effective manner, etc.
The MLIP package (short for Machine-Learning Interatomic Potentials) is available at \url{https://mlip.skoltech.ru/download/}.

\section{Introduction}

Machine-learning interatomic potentials have recently been a subject of research and now they are turning into a \emph{tool} of research.
Interatomic potentials (or force-fields) are models predicting potential energy (together with its derivatives) of interaction of a variable-size atomic system as a function of atomic positions and types.
Machine-learning potentials differ from classical, empirical potentials such as EAM or REAX-FF by a flexible functional form that can be systematically improved to approximate an arbitrary quantum-mechanical interaction subject to some assumptions like locality of interaction or smoothness of the underlying potential energy surface.
For about a decade, the focus of research has been on feasibility and properties of such an approximation---its accuracy, efficiency, and transferability depending on the chemical composition, nature of bonding, or level of quantum-mechanical theory being approximated.
In contrast, this manuscript describes how to practically apply the moment tensor potentials to seamlessly accelerate quantum-mechanical calculations, and the MLIP package (short for Machine-Learning Interatomic Potentials) implementing this tool.

The concept of machine-learning potentials was pioneered in 2007 by Behler and Parrinello \cite{behler2007-NNP}, who, motivated by the success of approximating potential energy surfaces of small molecules with neural networks, proposed the use of neural networks as a functional form of interatomic potentials, going beyond the small molecular systems by employing the locality of interaction.
The structure of such potentials consists of two parts: descriptors---usually two- and three-body ones---whose role is to describe local atomic environments accounting for all the physical symmetries of interaction, and a regressor---a function that maps the descriptors onto the interaction energy.
Neural network has been the first and are probably the most popular form of regressor for machine-learning potentials \cite{artrith2012-active-learning,behler2016-review,dolgirev2016-ML,smith2017-ANI,zhang2018-E-Car-deep-mlip,gastegger2015-high-dim-nn-organic,pun2019-mishin-ann}.
A related but slightly different class of potentials are the message-passing potentials that are based on neural networks, but go beyond the local descriptors of atomic environments \cite{schutt2017-schnet,lubbers2018-DNN}.

The other form of regressor proposed in 2010 was Gaussian processes used in the Gaussian Approximation Potentials (GAP) \cite{bartok2010-GAP,szlachta2014-tungsten,deringer2016-GAP-carbon,grisafi2018-ML-tensorial-properties,jinnouchi2019-on-the-fly}.
When used with the smooth overlap of atomic positions (SOAP) kernel \cite{bartok2013-soap,szlachta2014-tungsten}, they can provably approximate an arbitrary local many-body interaction of atoms, in contrast to the neural network potentials employed on top of two- and three-body descriptors \cite{pozdnyakov2020-completeness}\footnote{Strictly speaking, \cite{pozdnyakov2020-completeness} proves that \emph{not every} partitioning of the total energy into local contributions can be approximated as a function of two- and three-body descriptors. The extent to which this affects the accuracy of approximating the \emph{best} partitioning is yet to be investigated.}.
An alternative application of Gaussian processes was recently proposed \cite{vandermause2020-bayesian-on-the-fly} for explicit three-body potentials.

Probably the third largest class of interatomic potentials is based on linear regression with a set of basis functions. This includes the spectral neighbor analysis potential (SNAP) \cite{thompson2015-snap,wood2018-snap-quadratic} including the recent extension to multicomponent systems \cite{cusentino2020-snap-multielement}, and polynomial-based approaches \cite{drautz2019-ace,vanderoord2020-pip} including moment tensor potentials (MTP) \cite{shapeev2016-mtp}---the main focus of the present work.
Our extension of MTP to multicomponent systems \cite{gubaev2018-chemoinformatics,gubaev2019-alloys} goes beyond the linear regression, however, there is an alternative formulation of multicomponent MTP that stays linear \cite{lomaka2020-linear-multicomponent-MTP}.

There exist related approaches solving similar problems but falling outside the class of machine-learning potentials.
These are symbolic-regression potentials \cite{hernandez2019-symbolic-regression-potential}, non-conservative force fields \cite{botu2015-MLIP,devita2015-LOTF}, force-fields for a fixed-size molecular system (typically applied to one or two organic molecules) \cite{manzhos2015-finite-dim-neural,chmiela2017-finite-pes},
on-lattice potentials \cite{shapeev2017-lrp}, 
and cheminformatics-type models \cite{rupp2012-coulomb-matrix,snyder2012-ML-for-DFT,de2016-soap-cheminformatics,faber2017-cheminformatics,hansen2015-cheminformatics,huang2016-cheminformatics,gilmer2017-ceminformatics,schutt2017-DTNN,christensen2020-fchl-revisited}, see also a recent review \cite{lilienfeld2020-review-nature}.

\subsection*{Moment Tensor Potentials}

MTP has been first proposed as a single-component potential with linear dependence of the energy on the fitting parameters \cite{shapeev2016-mtp}.
The basis in which the potential was expanded is polynomial-like (adapted so that instead of exhibiting polynomial growth at large interatomic distances the basis functions stop feeling atoms that left the finite cutoff sphere).
This basis is very similar to the one of atomic cluster expansion (ACE), \cite{drautz2019-ace} and related to the permutation-invariant polynomial (PIP) basis \cite{vanderoord2020-pip}.
An active learning algorithm was proposed in \cite{podryabinkin2017-AL} based on which point defect diffusion \cite{novoselov2019-diffusion} and lattice dynamics \cite{ladygin2020-lattice-dynamics} of a number of mono-component crystals were studied.
Later MTP was generalized to multiple components by introducing nonlinear dependence on some of the parameters---first in the context of cheminformatics \cite{gubaev2018-chemoinformatics} and later as an interatomic potential \cite{gubaev2019-alloys}.
Moment tensor potential was applied to
predicting stable crystalline structures of single-components \cite{podryabinkin2019-uspex}
and convex hulls of stable alloy structures \cite{gubaev2019-alloys},
simulating lattice dynamics \cite{ladygin2020-lattice-dynamics},
calculating free energy of high-entropy alloy \cite{grabowski2019-hea},
calculating mechanical properties of a medium-entropy alloy \cite{jafary2019-hea},
computing lattice thermal conductivity of complex materials \cite{korotaev2019-conductivity},
studying diffusion of point defects in crystals \cite{novoselov2019-diffusion},
modeling covalent and ionic bonding \cite{novikov2019-SiO2},
computing molecular reaction rates \cite{novikov2018-rpmd,novikov2019-rpmd-SH2},
and studying a variety of properties of two-dimensional materials \cite{mortazavi2020-conductivity,mortazavi2020-phonon,raeisi2020-bohayra-diamanes,mortazavi20200-CN-nanosheets}, including enabling multiscale calculations of heat conductivity of polycrystalline materials \cite{mortazavi2020-multiscale} which are otherwise hard to carry out with classical, pre-Big Data modeling approaches.

MTPs, like SOAP-GAP, are not based on solely two- and three-body descriptors and can provably approximate an arbitrary local interaction, and so are ACE \cite{drautz2019-ace} and PIP \cite{vanderoord2020-pip}.
Probably because of this MTP together with GAP showed excellent accuracy in recent cheminformatics benchmark test \cite{nyshadham2019-comparison} and interatomic potential test \cite{zuo2020-benchmark}; in the latter MTP showed also a very good balance between accuracy and computational efficiency when compared against other machine-learning potentials.

\subsection*{Active learning}

A crucial, and often time-consuming part is the construction of the training set.
Traditionally, the training set is constructed through laborious trial-and-error iterations in each of which a researcher manually assesses the performance of the trained potential and tries to understand how to construct configurations for the new training set to avoid the undesirable behavior of the potential on the next iteration.
Active learning is a machine-learning technique allowing one to entrust these training set refinement iterations to a computer, thus completely automating the training set construction.

The ideas of active learning during atomistic simulations come from the concept of learning on-the-fly \cite{devita1997-on-the-fly,csanyi2004-on-the-fly} which can be retrospectively described as a learning-and-forgetting scheme.
In this scheme the training set would consist of some of the last configurations of the molecular dynamics trajectory.
A learning-and-remembering scheme was first proposed in \cite{li2015-on-the-fly}, however, the sampling was uniform---a fixed number of time steps were skipped before adding a configuration into the training set, which otherwise grows indefinitely.

For a more sophisticated strategy one needs an indicator of an error that a machine-learning model commits when trying to predict the energy and derivatives of a configuration \emph{without} making a quantum-mechanical calculation.
Such an indicator in the field of machine learning is called a \emph{query strategy} \cite{settles2012-AL}.
Such a strategy was first implemented by Artrith and Behler in \cite{artrith2012-active-learning} for the neural network potential.
In their work they used an ensemble of independently trained neural networks to ``vote'' for the predicted energy.
The deviation between different neural networks was taken as the sought indicator of the error.
It was then used to conduct a molecular dynamics simulation and adding to the training set those configurations on which different neural networks significantly disagreed.
This query strategy is called the \emph{query by committee}.
It is the algorithm of choice for other neural-network-based potentials as well \cite{smith2018-active-learning,zhang2019-e-active-learning}.
An exception is the work \cite{botu2015-adaptive} where the authors train two models: one predicting the energy and another predicting uncertainty.

Gaussian process-based potentials have another natural query strategy---predictive variance \cite{jinnouchi2019-on-the-fly,vandermause2020-bayesian-on-the-fly}.
Interestingly, for the original Gaussian approximation potentials other sampling criteria has been proposed  \cite{bernstein2019-denovo,sivaraman2019-gap-AL-weird}.

We, for the moment tensor potentials, employ a special form of what is known as the D-optimality criterion \cite{settles2012-AL}.
It employs a geometric criterion based on the so-called \emph{extrapolation grade}---a quantity characterizing the extent to which a given configuration is extrapolative with respect to those in the training set.
This algorithm was proposed for linearly parametrized, single-component MTPs in \cite{podryabinkin2017-AL} and generalized to nonlinear MTPs in \cite{gubaev2018-chemoinformatics,gubaev2019-alloys}.
The details of our algorithm will be given in Section \ref{sec:AL}.

\subsection*{Structure of this manuscript}

In this manuscript we focus on the methodology of applying MTPs and active learning to performing atomistic simulations.
In Section \ref{sec:theory} we describe our formulation of the moment tensors potentials and active learning.
Section \ref{sec:mlip-package} gives a brief overview of the MLIP package implementing MTPs and active learning, which is detailed in \cite{suppinfo-manual}.
Then, Sections \ref{sec:example-1}--\ref{sec:example-3} describe three practical examples of the use of MTPs and active learning to perform atomistic calculations, with the focus on the particular steps of the calculation.
The files that are referenced in Sections \ref{sec:example-1}--\ref{sec:example-3} are available in our supplemental information \cite{suppinfo-examples}, as well as in the \verb|doc/examples/| folder of the MLIP package \cite{mlip-download}.
The description of and references to the MLIP code is kept to minimum, yet retained for the purpose of easier preproduction of the described results.

\section{Theory}\label{sec:theory}

\subsection{Moment Tensor Potential} \label{sec:MTP}

Moment Tensor Potentials belong to the class of machine-learning potentials implemented in the MLIP package.
These potentials represent the energy of an atomic configuration \verb|cfg| as a sum of contributions of local atomic environments of each atom.
The atomic environment, or neighborhood, $\mathfrak{n}_i$ of the $i$th atom is comprised of its atomic type $z_i$, the atomic type of its neighbors, $z_j$ and positions of the neighbors relative to the $i$th atom, $\bm{r}_{ij}$.
The potential energy of interatomic interaction, $E^{\rm mtp}$, is thus
\begin{equation} \label{eq:E_MTP}
E^{\rm mtp}(\cfg) = \sum_{i=1}^n V(\mathfrak{n}_i).
\end{equation}
The function $V$ is linearly expanded through a set of basis functions $B_{\alpha}$:
\begin{equation} \label{eq:V_MTP}
V(\mathfrak{n}_i) = \sum \limits_{\alpha} \xi_{\alpha} B_{\alpha}(\mathfrak{n}_i),
\end{equation}
where ${\bm \xi} = \{ \xi_{\alpha} \}$ are parameters to be found by fitting them to the training set.

\begin{figure}[ht]
	\centering
	\includegraphics{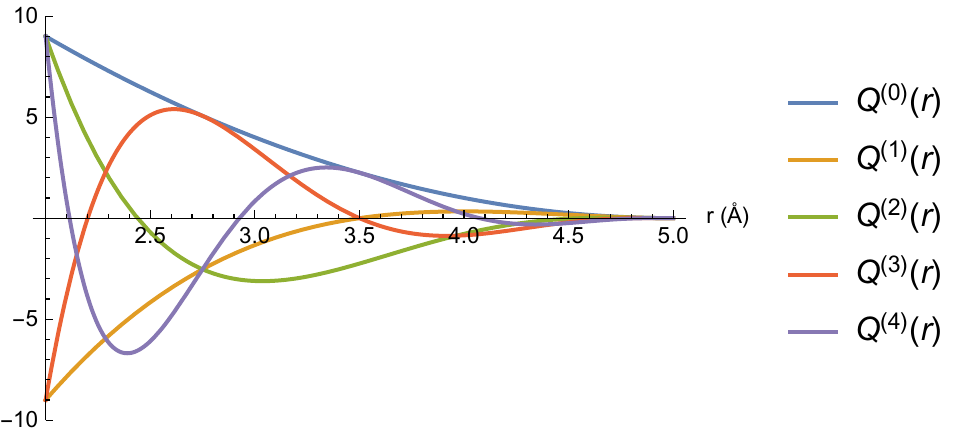}
	\caption{Radial basis functions $Q^{(\beta)}(r)$ as defined in \eqref{eq:radial-basis}.
		Plotted for $R_{\rm min}=2\angstrom$, $R_{\rm cut}=5\angstrom$, $0 \leq \beta \leq 4$.
	}
	\label{fig:radial-functions}
\end{figure}

To define the functional form of the basis functions $B_{\alpha}$, we introduce the moment tensor descriptors, or simply \emph{moments}: 
\[
M_{\mu,\nu}({\mathfrak{n}}_i)=\sum_{j} f_{\mu}(|r_{ij}|,z_i,z_j) \underbrace {\bm{r}_{ij}\otimes...\otimes \bm{r}_{ij}}_\text{$\nu$ times};
\]
the notation is detailed below.
These descriptors consist of the radial and angular part. The radial part has the form
\begin{equation}\label{eq:f_mu}
f_{\mu}(|r_{ij}|,z_i,z_j) = \sum_{\beta=1}^{N_Q} c^{(\beta)}_{\mu, z_i, z_j} Q^{(\beta)} (|r_{ij}|),
\end{equation}
where ${\bm c} = \big\{ c^{(\beta)}_{\mu, z_i, z_j} \big\}$ is the set of ``radial'' parameters.
We call the functions $Q^{(\beta)} (|r_{ij}|)$ the radial basis functions:
\begin{equation} \label{eq:radial-basis}
\displaystyle
Q^{(\beta)}(|r_{ij}|) =\begin{cases}
\varphi^{(\beta)} (|r_{ij}|) (R_{\rm cut} - |r_{ij}|)^2 & |r_{ij}|<R_{\rm cut} \\
0 & |r_{ij}| \geq R_{\rm cut}.
\end{cases}
\end{equation}
Here $\varphi^{(\beta)}$ are polynomial functions (e.g., Chebyshev polynomials) on the interval $[R_{\rm min}, R_{\rm cut}]$, where $R_{\rm min}$ is the minimal distance between atoms in the system investigated, $R_{\rm cut}$ is the cutoff radius which is introduced to ensure a smooth behavior of MTP when atoms leave or enter the interaction neighborhood.
An illustration of the radial basis functions is given in Figure \ref{fig:radial-functions}.

The angular part $\underbrace {\bm{r}_{ij}\otimes...\otimes \bm{r}_{ij}}_\text{$\nu$ times}$ contains angular information about the neighborhood $\mathfrak{n}_i$. Here the symbol $``\otimes"$ is the outer product of vectors, and, thus, the angular part is the tensor of rank $\nu$. E.g., for $\nu = 0$ the angular part is a scalar and simply equals $1$, if $\nu = 1$ the angular part is the vector $\bm{r}_{ij} = (x_{ij}, y_{ij}, z_{ij})$ pointing from atom $i$ to atom $j$, if $\nu = 2$ the angular part has the form of the matrix:
\[
\bm{r}_{ij} \otimes \bm{r}_{ij} = \left(\begin{matrix}
x_{ij}^2 & x_{ij} y_{ij} & x_{ij} z_{ij} \\
y_{ij} x_{ij}& y_{ij}^2 & y_{ij} z_{ij} \\
z_{ij} x_{ij} & z_{ij} y_{ij} & z_{ij}^2 \\
\end{matrix}\right).
\]

In order to construct the basis functions $B_{\alpha}$ we define the so-called \emph{level} of moments: 
\begin{equation} \label{eq:LevelMTD}
{\rm lev} M_{\mu,\nu} = 2 + 4 \mu + \nu,
\end{equation}
for example
${\rm lev} M_{0,1} = 3$, 
${\rm lev} M_{1,1} = 7$,
${\rm lev} M_{0,2} = 4$,
${\rm lev} M_{0,0} = 2$.
The coefficients 2, 4, and 1 in \eqref{eq:LevelMTD} were empirically found to be optimal on a number of tests done in \cite{gubaev2019-alloys} and are fixed in the MLIP package.
The level of multiplication, or more generally, contractions of a number of moments is defined by adding the levels, for example
\begin{align*}
{\rm lev} M_{1,0}^2 &= 12, &
{\rm lev} M_{0,0}^4 &= 8, &
{\rm lev} M_{2,0}^3 &=30, \\
{\rm lev} (M_{1,1} \cdot M_{0,1}) &= 10, &
{\rm lev} (M_{1,2} : M_{0,2}) &= 12,&
{\rm lev} ((M_{0,3}M_{0,2}) \cdot M_{0,1}) &= 12,
\end{align*}
where $``\cdot"$ is the dot product of two vectors, $``:"$ is the Frobenius product of two matrices.

As could be seen from these examples, non-scalar moments could yield scalars upon their contraction.
All such contractions of one or more moments are, by definition, MTP basis functions $B_{\alpha}$.
Each basis function is, by its definition, invariant to atomic permutations, rotations, and reflections.
Finally, to define a particular functional form of MTP, we choose the maximum level, $\rm {lev}_{\rm{max}}$, and include all the basis functions whose level is less or equal than that: ${\rm lev} B_{\alpha} \leq \rm {lev}_{\rm{max}}$.
For example, if we choose $\rm {lev}_{\rm{max}} = 8$ then we have nine basis functions $B_{\alpha}$:
\begin{equation} \label{eq:MTPbasisLevel8}
\begin{array}{l@{~~}l}
B_1=M_{0,0},& {\rm lev} B_1 = 2; \\
B_2=M_{1,0},& {\rm lev} B_2 = 6; \\
B_3=M_{0,0}^2,& {\rm lev} B_3 = 4; \\
B_4=M_{0,1} \cdot M_{0,1},& {\rm lev} B_4 = 6; \\
B_5=M_{0,2} : M_{0,2},& {\rm lev} B_5 = 8; \\
B_6=M_{0,0}M_{1,0},& {\rm lev} B_6 = 8; \\
B_7=M_{0,0}^3,& {\rm lev} B_7 = 6; \\
B_8=M_{0,0} (M_{0,1} \cdot M_{0,1}),& {\rm lev} B_8 = 8; \\
B_9=M_{0,0}^4,& {\rm lev} B_9 = 8.
\end{array}
\end{equation}

The radial parameters ${\bm c}$ from \eqref{eq:f_mu} together with ${\bm \xi}$ from \eqref{eq:V_MTP} comprise the set of MTP parameters ${\bm \theta} = \{ \bm \xi, \bm c \}$ that are found by fitting to the training set as described in the next section.
Thus, the energy as predicted by MTP will be denoted as $E^{\rm mtp}(\cfg;{\bm \theta})$ when we want to emphasize the dependence on ${\bm \theta}$.

Thus, the functional form of MTP is determined by the two numbers---level of MTP ${\rm lev}_{\rm{max}}$ and size of the radial basis $N_Q$ defined in \eqref{eq:f_mu}.
The number of basis functions (and the number of the corresponding parameters $\bm \xi$) grows exponentially with ${\rm lev}_{\rm{max}}$, while the number of radial functions $f_\mu$ (and the number of the corresponding parameters $\bm c$) grows as $O(N_Q\, {\rm lev}_{\rm{max}})$.
The hyperparameters ${\rm lev}_{\rm{max}}$ and $N_Q$ should be chosen, for a particular application, to achieve the desired balance between the accuracy of MTP, computational efficiency of MTP, and number of the required quantum-mechanical calculations, the latter is usually proportional to the total number of free parameters in MTP.

\subsection{Training on a quantum-mechanical database}

Let the training set contain the configurations $\cfg_k$, $k=1,\ldots,K$, with known quantum-mechanical energies $E^{\rm qm}(\cfg_k)$, forces ${\bm f}_i^{\rm qm}(\cfg_k)$, and stress tensors $\sigma^{{\rm qm}}(\cfg_k)$, where index $i$ goes through the atoms in configuration $\cfg_k$.
The passive learning (fitting, training) of MTP consists of finding the parameters $\bm \theta$ during solving the machine-learning (optimization) problem:
\begin{equation} \label{Fitting}
\begin{array}{c}
\displaystyle
\sum \limits_{k=1}^K \Bigl[
	w_{\rm e} \left(E^{\rm mtp} (\cfg_k; {\bm {\theta}}) - E^{\rm qm}(\cfg_k) \right)^2
	+
	w_{\rm f} \sum_{i=1}^{N_k} \left| {\bm f}^{\rm mtp}_i(\cfg_k; {\bm {\theta}}) - {\bm f}^{\rm qm}_i(\cfg_k) \right|^2 
\\ \displaystyle
	+
	w_{\rm s} \big|\sigma^{\rm mtp}(\cfg_k; \bm {\theta}) - \sigma^{{\rm qm}}(\cfg_k)\big|^2 \Bigr] \to \min\limits_{\bm \theta},
\end{array}
\end{equation}
where $N_k$ is the number of atoms in the $k$-th configuration, $w_{\rm e}$, $w_{\rm f}$, and $w_{\rm s}$ are non-negative weights expressing the importance of energies, forces, and stresses in the optimization problem.
Here for a stress tensor $\sigma$ by $|\sigma|^2$ we mean the Frobenius norm, $|\sigma|^2 = \sum_{\alpha,\beta=1}^3 |\sigma_{\alpha\beta}|^2$.

After training we can measure the root-mean-square errors in energy, forces, and stresses as
\begin{align} \label{eq:E_RMSE}
{\rm RMSE}(E)^2 &= \dfrac{1}{K} \sum \limits_{k=1}^K \left( \dfrac{E^{\rm mtp} (\cfg_k; {\bm {\theta}})}{N^{(k)}} - \dfrac{E^{\rm qm}(\cfg_k)}{N^{(k)}} \right)^2,
\\ \label{eq:F_RMSE}
{\rm RMSE}({\bm f})^2 &= \dfrac{1}{K} \sum \limits_{k=1}^K \dfrac{1}{3 N^{(k)}} \sum_{i=1}^{N_k} \left| {\bm f}^{\rm mtp}_i(\cfg_k; {\bm {\theta}}) - {\bm f}^{\rm qm}_i(\cfg_k) \right|^2,
\\ \label{eq:S_RMSE}
{\rm RMSE}(\sigma)^2 &= \dfrac{1}{K} \sum \limits_{k=1}^K \,\dfrac{1}{9}\,\big|\sigma^{\rm mtp}(\cfg_k; \bm {\theta}) - \sigma^{{\rm qm}}(\cfg_k)\big|^2.
\end{align} 
The mean absolute errors (MAE) and maximum absolute errors are defined accordingly.
It is often a good idea to set aside a validation (or test) set of configurations on which to measure the errors according to the formulae \eqref{eq:E_RMSE}--\eqref{eq:S_RMSE}---this would give an unbiased, unaffected by overfitting, estimate of the error of the potential.

Depending on the variety and size of configurations in the training set we may consider different weighting of configurations depending on the number of atoms, $N_k$, in the optimization problem.
The main criterion is that the same configuration with a larger unit cell should have the same relative contributions of macroscopic properties (energy and stresses), and microscopic properties (forces). Thus, if the training set consists only of configurations of not more than tens of atoms and in which unit cell (and, also, stresses) does not play an important role (e.g., organic molecules), we may find MTP parameters by solving the problem \eqref{Fitting} with $w_{\rm s} = 0$, and without any scaling by $N_k$.
We tag the weighting given by \eqref{Fitting} as ``molecules''.

If we have configurations of different size in the training set (the reader may think of different supercells of the same structure) but we want all the structures to have the same weight in the training set, irrespective of the number of atoms that are used to represent it, then we use the following scaling (tagged as ``structures''):
\begin{equation} \label{eq:train-structures}
\begin{array}{c}
\displaystyle
\sum \limits_{k=1}^K \Bigl[ \dfrac{w_{\rm e}}{(N_k)^2} \left(E^{\rm mtp} (\cfg_k; {\bm {\theta}}) - E^{\rm qm}(\cfg_k) \right)^2 + \dfrac{w_{\rm f}}{N_k} \sum_{i=1}^{N_k} \left| {\bm f}^{\rm mtp}_i(\cfg_k; {\bm {\theta}}) - {\bm f}^{\rm qm}_i(\cfg_k) \right|^2 
\\
\displaystyle
+ \dfrac{w_{\rm s}}{(N_k)^2} \big|\sigma^{\rm mtp}(\cfg_k; \bm {\theta}) - \sigma^{{\rm qm}}(\cfg_k)\big|^2 \Bigr] \to \operatorname{min}.
\end{array}
\end{equation}

Finally, if we are studying thermal properties with molecular dynamics then we assume that in two configurations with different atoms each force vector has the same importance and the importance of the energy grows as $\sqrt{N_k}$, hence in this case we consider the following scaling (tagged as ``vibrations''):
\begin{equation} \label{eq:train-vibrations}
\begin{array}{c}
\displaystyle
\sum \limits_{k=1}^K \Bigl[ \dfrac{w_{\rm e}}{N_k} \left(E^{\rm mtp} (\cfg_k; {\bm {\theta}}) - E^{\rm qm}(\cfg_k) \right)^2 + w_{\rm f} \sum_{i=1}^{N_k} \left| {\bm f}^{\rm mtp}_i(\cfg_k; {\bm {\theta}}) - {\bm f}^{\rm qm}_i(\cfg_k) \right|^2 
\\
\displaystyle
+ \dfrac{w_{\rm s}}{N_k} \big|\sigma^{\rm mtp}(\cfg_k; \bm {\theta}) - \sigma^{{\rm qm}}(\cfg_k)\big|^2 \Bigr] \to \operatorname{min}.
\end{array}
\end{equation}

We refer to the method described in this section as passive learning of MTP because here we generate our training set manually and MTP is not ``choosing'' what to train itself on.
Our MLIP code also allows one to select configurations for the training set automatically, using the so-called active learning algorithm, presented in the next section.

\subsection{Active learning: query strategy} \label{sec:AL}

The quality of a machine-learning potential is determined not only by a functional form (an efficient representation) but also by the quality of the training set.
A known drawback of machine-learning potentials is poor prediction for configurations that are far from the training set (see, e.g., \cite{szlachta2014-W-GAP}).
This could be described by saying that we should assemble the training set such that during the simulation (e.g., molecular dynamics) the potential is ``interpolating'' with respect to the training set when trying to make predictions for energy, forces, and stresses.
An extrapolation may lead to significant errors causing instability of atomistic simulation. For example, it is hard to expect an accurate simulation of a free surface with a potential fitted only on the bulk configurations.
The notion of interpolation and extrapolation can even be rationalized mathematically \cite{podryabinkin2017-AL}.

Traditionally, the training set for interatomic potentials is constructed manually through several loops of trial-and-error, manually assessing the quality of the potential.
However, with a formal definition of extrapolation it is possible to automate this procedure reducing months of manual work to hours of computer time.
Our definition of extrapolation is based on the D-optimality criterion \cite{podryabinkin2017-AL} postulating that a good training set is the one that corresponds to the maximial value of the determinant of the information matrix \cite{settles2012-AL}.
To be precise, we introduce the notion of extrapolation grade $\gamma(\cfg)$---a feature of a configuration (and a training set) that correlates with the prediction error but does not require ab initio information to be calculated prior to its evaluation; its precise definition is given below.

Once we have defined the extrapolation grade $\gamma(\cfg)$, we can formulate our {\it query strategy}: we collect all the configurations occurring in a simulation whose grade $\gamma$ is higher than a chosen threshold.
Such configurations are later computed with an ab initio model and added to the training set.
This dynamically ensures that during a simulation no significant extrapolation occurs.

The D-optimality criterion that MLIP is based on, is easiest to understand in the context of linear regression, i.e., in the case when MTP is parametrized only by linear parameters (${\bm \xi} = \{ \xi_1 \dots \xi_m\}$). According to \eqref{eq:E_MTP} and \eqref{eq:V_MTP} the energy of a configuration in this case can be expressed as   
\[
E^{\rm mtp}(\cfg; \bm \xi) = \sum_{i} \sum \limits_{\alpha=1}^m \xi_{\alpha} B_{\alpha}(\mathfrak{n}_i) = \sum \limits_{\alpha=1}^m \xi_{\alpha} \underbrace{\sum_{i} B_{\alpha}(\mathfrak{n}_i)}_{b_\alpha(\cfg)}.
\]
When fitting to the energy values, we need to solve an overdetermined system of $K$ linear equations on ${\bm \xi}$ with the matrix
\begin{equation}\label{eq:matrix-B}
\mathsf B = 
\begin{pmatrix}
b_1(\cfg_1) & \ldots & b_m(\cfg_1) \\
\vdots & \ddots & \vdots \\
b_1(\cfg_K) & \ldots & b_m(\cfg_K) \\
\end{pmatrix}
.
\end{equation}
Each equation of this system is produced by a certain configuration.
In our version of the D-optimality criterion we select $m$ configurations that yield a set of the most linearly independent equations in the sense that the corresponding $m\times m$ submatrix $\mathsf A$ has the maximal modulus of determinant, $|\rm det(\mathsf A)|$ (maximal volume).
We call the $m$ selected configurations the \textit{active set} and in the MLIP code the active set together with $\mathsf A$ and ${\mathsf A}^{-1}$ is called the \textit{active learning state}.
Note that the D-optimal active set corresponds to the most ``extreme'' and diverse configurations from the point of view of MTP.

A practical way to construct an active set from a pool of configuration is provided by the Maxvol algorithm for finding the submatrix of maximal volume in a tall matrix \cite{goreinov2010-maxvol}. Assume that we already have some active set of configurations and the corresponding rows of the square matrix $\mathsf A$.
We can then test whether a candidate configuration can increase $|\rm det(\mathsf A)|$ by replacing another configuration from the current active set. For this purpose we compute 
\begin{align*}
\gamma(\cfg) &=\max_{1\leq j \leq m} |c_j|,\qquad\text{where}
\\
\begin{pmatrix}
c_1 & \ldots & c_m
\end{pmatrix} &=
\begin{pmatrix}
b_1 & \ldots & b_m
\end{pmatrix}
 \mathsf A^{-1}.
\end{align*}
Thus, if $\gamma(\cfg)>1$ then $|\rm det(A)|$ could be increased.
In this case $(b_1\, \ldots\, b_m)$ should, according to the Maxvol algorithm, replace the row with index 
\[
k={\rm argmax}_{1\leq j \leq m} |c_j|
\]
of the matrix $\mathsf A$, and the active set should be hence updated. 

Note that if the active set and the training set are the same, then the parameters can be found as 
\[
{\bm \xi} = \big(E^{\rm qm}(\cfg_1) \, \ldots\, E^{\rm qm}(\cfg_m) \big) \mathsf A^{-1}
\]
and the energy of any configuration can be expressed as 
\[
E^{\rm mtp}(\cfg) = \sum_{j=1}^m c_j  E^{\rm qm}(\cfg_j).
\]
This mathematical formula allows us to formally say that MTP extrapolates if $\gamma(\cfg)>1$, and interpolate otherwise. We hence interpret $\gamma$ as the \textit{extrapolation grade}. 

\begin{figure}[ht]
	\centering
	\includegraphics[width=\textwidth]{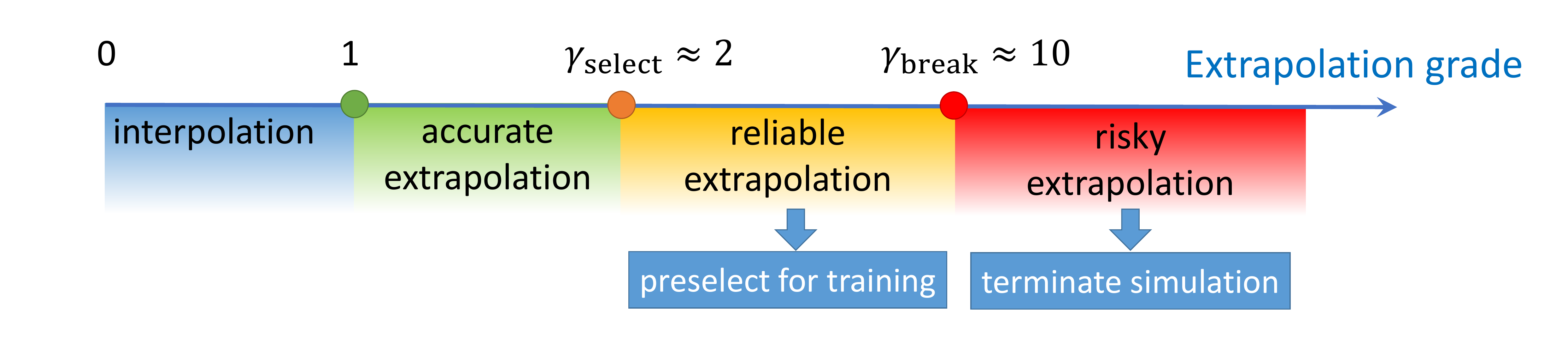}
	\caption{Classification of extrapolation grades. A configuration with the grade less than $\gamma_{\rm select}$ does not trigger any active learning actions, a configuration whose grade is between $\gamma_{\rm select}$ and $\gamma_{\rm break}$ are considered reliable yet useful for extension of the training set, but those with the grade larger than $\gamma_{\rm break}$ are risky and trigger termination of the simulation.}
	\label{Fig:thresholds}
\end{figure}

In practical simulations with MLIP we use the so-called two-threshold scheme illustrated in Figure \ref{Fig:thresholds}.
In this scheme we choose two thresholds: $\gamma_{\rm select}$---threshold beyond which we select a configuration for training, and $\gamma_{\rm break}$---threshold beyond which we terminate the simulation.
The rationale for this is the following.
If $\gamma_{\rm select}$ is not too high (typically around 2) then for configurations $\cfg$ with $\gamma(\cfg) < \gamma_{\rm select}$ the error of prediction is usually not significantly higher than for interpolative configurations.
When $\gamma_{\rm select} < \gamma(\cfg) < \gamma_{\rm break} \approx 10$ the error of predictions may be significantly higher, however, the simulation still remains reliable and does not have to be terminated.
The values $1.1 \lesssim \gamma_{\rm select} \lesssim 5$ and $3 \lesssim \gamma_{\rm break} \lesssim 20$ appear to be universal for any atomistic system that has been tried so far, however, the researchers are advised to do their own testing.

To generalize the D-optimality criterion to nonlinearly parametrized MTP, assume that the values of the parameters, $\bar{\bm \theta}$, are already near the optimal ones and we hence linearize the energies ${\bf E}^{\rm mtp}$ with respect to the parameters.
The matrix $\mathsf B$ in this case is a tall Jacobi matrix
\[
\mathsf B = \begin{pmatrix}
\frac{\partial}{\partial \theta_1} E^{\rm mtp}\big(\cfg_1; \bar{\bm\theta}\big) & \ldots & \frac{\partial}{\partial \theta_m} E^{\rm mtp}\big(\cfg_1; \bar{\bm\theta}\big) \\
\vdots & \ddots & \vdots \\
\frac{\partial}{\partial \theta_1} E^{\rm mtp}\big(\cfg_K; \bar{\bm\theta}\big) & \ldots & \frac{\partial}{\partial \theta_m} E^{\rm mtp}\big(\cfg_K; \bar{\bm\theta}\big) \\
\end{pmatrix},
\]
where each row corresponds to a particular configuration from the training set.
The matrix $\mathsf A$ is represented in the analogous manner, and other details of the active learning algorithm remain the same.


\subsection{Active learning bootstrapping iterations} \label{sec:LOTF}

As discussed in the previous section, a good way of assembling a training set is by sampling configurations directly from the atomistic simulation we are planning to conduct.
This is often called learning on-the-fly, following pre-machine-learning algorithms \cite{csanyi2004-on-the-fly}.
However, in the simple learning-on-the-fly strategy there are issues, for instance, with energy conservations after refitting the potential.
To address that, we use a bootstrapping technique in which we terminate the simulation, retrain a potential, and restart a simulation.
This way, no change in the underlying potential energy surface during a single simulation is committed.

\begin{figure}[ht]
	\centering
	\includegraphics[width=\textwidth]{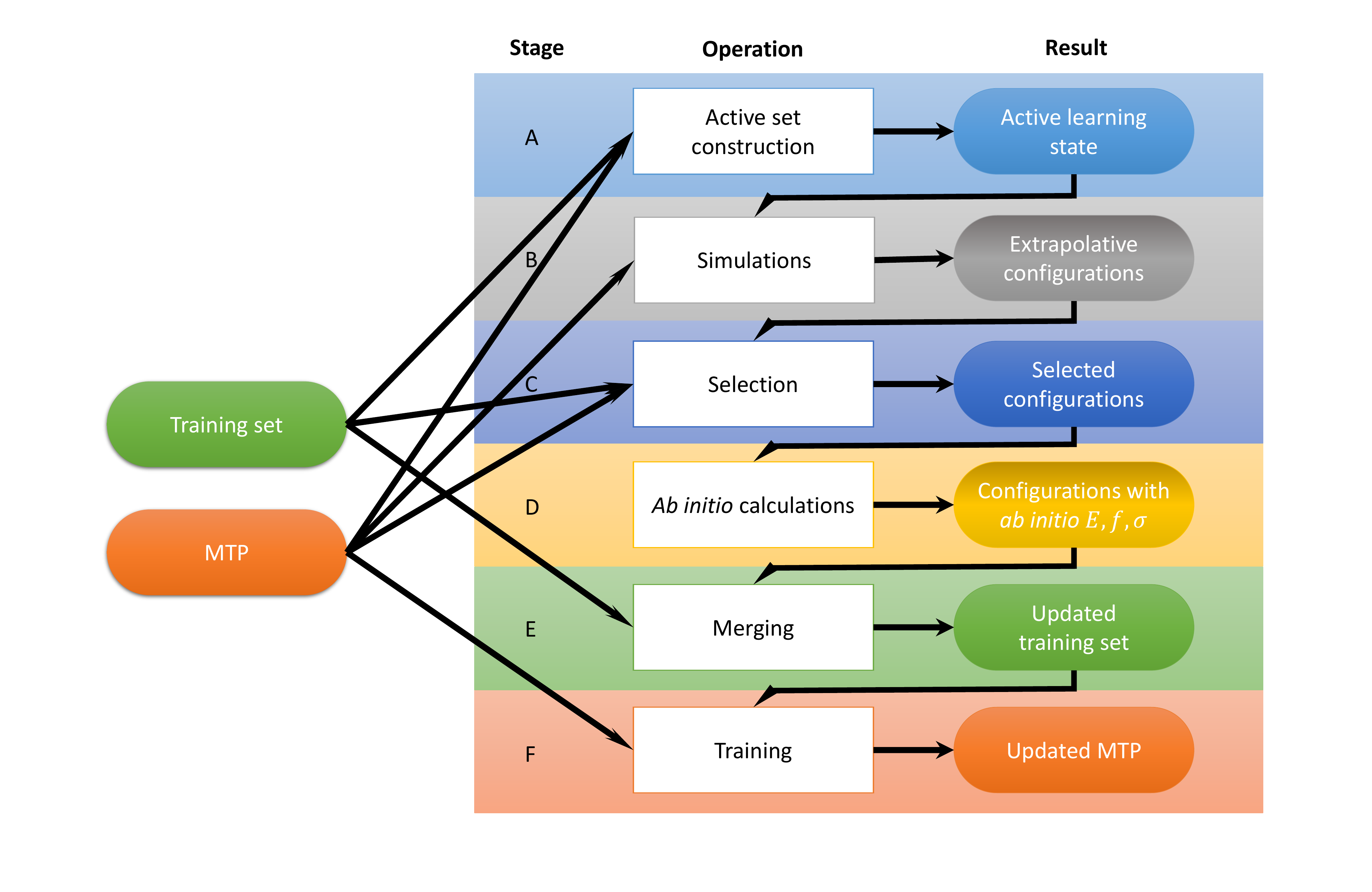}
	\caption{Scheme of active learning bootstrapping iterations.}
	\label{fig:lotf}
\end{figure}

Thus, one iteration of an active learning algorithm consists of the following steps (see Fig. \ref{fig:lotf}).
\begin{itemize}
	\item[A.] On the first step, the active set is selected among all configurations from the training set and the active learning state is formed.

	\item[B.] Run the simulation with the current potential and active selection of the extrapolative configurations ($\gamma(\cfg) > \gamma_{\rm select}$). The simulation is running until its successful completion or until $\gamma(\cfg) \geq \gamma_{\rm break}$.
	
	\item[C.] If a simulation stopped after exceeding the maximum allowed extrapolation grade ($\gamma(\cfg) \ge \gamma_{\rm break}$), an update of the active set has to be performed.
	To that end, the Maxvol algorithm is used to select the new configurations that will be appended to the training set among all extrapolative configurations sampled at the first step.
	
	\item[D.] The selected configurations are calculated with the ab initio model.
	
	\item[E.] Next the selected configurations with the ab initio the energy, forces, and stresses are appended to the trained set.
	
	\item[F.] The MTP is retrained.
\end{itemize} 

Each iteration of this scheme extends the training region and improves the stability of the MTP (i.e., simulation runs longer without termination by extrapolation). The iterations proceed until the simulation is finished without exceeding the critical value of extrapolation $\gamma_{\rm break}$.

\section{MLIP Package}\label{sec:mlip-package}

MLIP is a software package implementing moment tensor potentials.
It is distributed for free for non-commercial purposes and can be obtained at \cite{mlip-download}.
We briefly describe this software package here, and in detail in \cite{suppinfo-manual}.

The package can be compiled into the library providing interface between MLIP and other packages, most notably LAMMPS \cite{plimpton1993-lammps}, and \texttt{mlp} binary that provides basic operations including:
\begin{description}[style=nextline,leftmargin=\widthof{\ttfamily select-addXXX|},font=\normalfont\ttfamily]
\item[\texttt{convert-cfg}]
	converting VASP or LAMMPS input/output files to the internal \texttt{.cfg} format of atomic configuration
\item[\texttt{train}]
	training MTPs on \texttt{.cfg} datasets,
\item[\texttt{mindist}] 
    computes the minimal distances in configurations,
\item[\texttt{calc-efs}]
	using them to evaluate energy, forces, and stresses on a database,
\item[\texttt{calc-grade}]
	computing extrapolation grades and creating an active learning state for active-learning simulations,
\item[\texttt{select-add}]
	selecting a limited number of configurations from a large set of extrapolative configurations to be added to the training set (the name of command comes from two actions: selecting and adding),
\item[\texttt{relax}]
	use the internal structure relaxation algorithm to relax (i.e., minimize the potential energy of) configurations from a file.
\end{description}

A list and a short description of the commands can be obtained by executing
\begin{verbatim}
mlp list
\end{verbatim}
and 
\begin{verbatim}
mlp help <command>
\end{verbatim}
The commands \texttt{train} and \texttt{relax} work with MPI parallelism, the rest of the commands are serial.
The package contains untrained potentials defining MTPs of level 2, 4, \ldots, 28.

When called from what we call atomistic simulation ``drivers''---external codes or the \texttt{relax} command---the behavior of MLIP is controlled by the MLIP settings file, most often named \texttt{mlip.ini}, described in \cite{suppinfo-manual}.
There one can specify what MTP file to use, whether active learning should be switched on, the values of extrapolation thresholds, etc.

\section{Example 1: Elastic Constants of Molybdenum}\label{sec:example-1}

In this example we demonstrate how to passively train an MTP, calculate training and validation errors, and calculate energy/volume curve and elastic constants using bcc-Mo as an example.
The files referenced in this section are available at \cite{suppinfo-examples}.

\subsection{Training and Validation of MTP}

We choose the functional form of MTP of level 16 (the file \verb|untrained_mtps/16.mtp| of \cite{mlip-download}), eight radial basis functions, and set $R_{\rm cut} = 5.2\,\angstrom$ and $R_{\rm min} =  1.9\,\angstrom$.
We fit an ensemble of five MTPs in order to estimate the uncertainty of predictions---we will show that our estimated uncertainty reliably predicts the error of MTP as compared to DFT.
We take the training and test sets from \cite{zuo2020-benchmark} available at the public git repository\footnote{\texttt{https://github.com/materialsvirtuallab/mlearn}}.
We filter out configurations with the minimal distance between atoms smaller than 1.9 $\angstrom$.
The filtering and checking was facilitated by the \verb|mlp mindist| command.
We further recompute both sets of configurations with the VASP package \cite{VASP1,VASP3,VASP4} with slightly higher DFT convergence parameters, namely, the energy cutoff \verb|ENCUT| = 400 eV and the $\Gamma$-centered k-point mesh with \verb|KSPACING| = 0.114.
To fit an MTP we run the following command:
{\small\begin{verbatim}
mlp train init.mtp train.cfg --trained-pot-name=pot.mtp --valid-cfgs=test.cfg
\end{verbatim}}
The \verb|init.mtp| file does not include the fitting parameters, so they are initialized randomly in the beginning of training.
The file \verb|train.cfg| contains the training set in the MLIP format.
The \verb|valid-cfgs| option enables calculation of validation errors (i.e., the errors calculated on the test set) in the end of training, which otherwise could also be computed with the \verb|mlp calc-errors| command.
We have used the default values of the fitting weights $w_{\rm e} = 1$, $w_{\rm f} = 0.01$, and $w_{\rm s} = 0.001$ and the default fitting scaling \eqref{eq:train-vibrations}.
By repeating the training five times we obtain five potentials with the same functional form but different values of parameters due to different random initialization of the parameters.
We refer to these five MTPs as the ensemble of MTPs on which we compute the uncertainty of their predictions.
The average training and validation root-mean-square (RMS) errors, for the ensemble of MTPs and their uncertainty (2-sigma or 95\% confidence interval) are shown in Table \ref{tabl:TrainingValidationErrors}.
These errors are close to each other, and the uncertainty in the errors is small, indicating reliability of training.

\begin{table}
	\begin{center}
		\begin{tabular}{|c|c|c|c|} \cline{2-4}
			\multicolumn{1}{c|}{} & energy error & force error & stress error \\
			\multicolumn{1}{c|}{} & meV/atom & meV/\angstrom\ (\%) & GPa ($\%$) \\ \hline
			training   & 5.54 $\pm$ 2.10 & 175 $\pm$ 4 & 0.46 $\pm$ 0.04 \\ 
			& & (12.3 $\pm$ 0.4\%) & (4.4 $\pm$ 0.4\%) \\ \hline
			validation & 5.10 $\pm$ 1.12 & 180 $\pm$ 6 & 0.46 $\pm$ 0.09 \\ 
			& & (12.6 $\pm$ 0.4\%) & (4.4 $\pm$ 0.8\%) \\ \hline
		\end{tabular}
		\caption{\label{tabl:TrainingValidationErrors} Average training and validation errors for the ensemble of five MTPs and their uncertainty estimation with 95\% (i.e., 2-sigma) confidence interval. The uncertainty in the errors is small.}
	\end{center}
\end{table}

\subsection{Energy/Volume curve and elastic constants}

\begin{figure*} \begin{center}
	\includegraphics{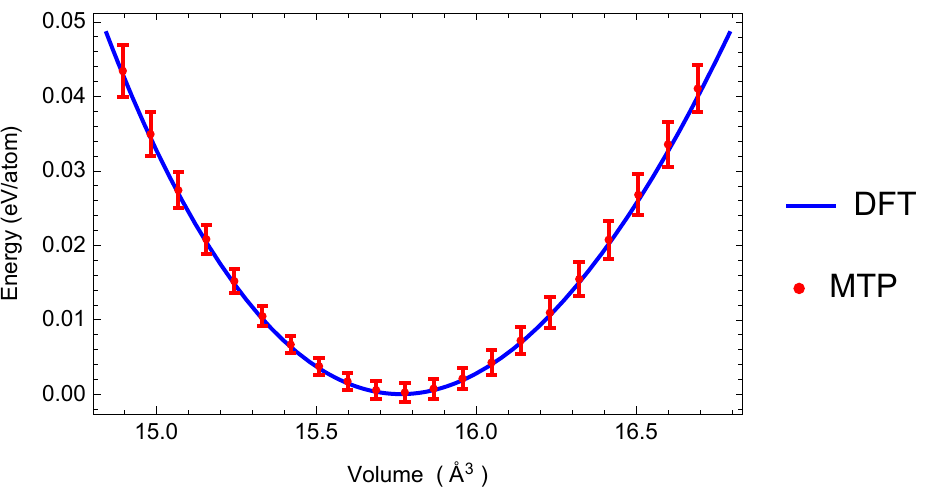}
\begin{center}
\caption{\label{Fig:EnergyVolumeCurve}
	Energy-volume curve for MTP and DFT.
	An ensemble of five MTPs was used to assess the uncertainty of prediction.
	The error bars correspond to the 95\% confidence interval (i.e., 2-sigma interval).
	One can see that the uncertainty is rather small---on the order of a meV/atom---and is indeed centered around the DFT values.
}
\end{center}
\end{center} \end{figure*}

For the calculation of elastic constants we first find the equilibrium lattice constant by finding the minimum of the energy/volume curve.
To that end, we generate the file \verb|deformed.cfg| with bcc-Mo configurations with compressed/stretched lattice constant.
We calculate the energies of these configurations by executing
\begin{verbatim}
mlp calc-efs pot.mtp deformed.cfg deformed_efs.cfg 
\end{verbatim}
The configurations with the energies calculated are written to \verb|deformed_efs.cfg|.
The energy volume/curve is shown in Figure \ref{Fig:EnergyVolumeCurve}.
The calculated MTP lattice constant together with its uncertainty is $3.15928 \pm 0.00097$ \angstrom\ (95\% confidence interval computed on the ensemble of five potentials).  
The reference DFT lattice constant is $3.15918$ \angstrom ---well within the MTP confidence interval.

We next calculate the elastic constants $C_{11}$, $C_{12}$, and $C_{44}$ of bcc-Mo using the finite difference method by applying $\pm 2\%$ strain to individual unit cell components.
We prepare the corresponding configurations in \verb|.cfg| files and evaluate their stresses with the \verb|calc-efs| command for MTP, and VASP calculations for DFT, with the same parameters as described above.

The elastic constants calculated with the ensemble of MTPs and using DFT are given in Table \ref{tabl:ElasticConstants}. As could be seen, the uncertainty of the calculation is relatively small, the MTP and DFT constants are close to each other, and the DFT constants fall well within the 95\% confidence interval of MTP.

We cross-check our finite-difference results using the LAMMPS script \verb|in.elastic|, adopted from the corresponding script from the \verb|examples/ELASTIC/| folder of LAMMPS.
To run the script with MTP, one must simply declare the MTP \verb|pair_style| potential in the script:
\begin{verbatim}
pair_style  mlip mlip.ini
pair_coeff  * *
\end{verbatim}
where \verb|mlip.ini| is the file with the MLIP settings indicating that \verb|pot.mtp| should be used.
The script directly yields the elastic constants of bcc-Mo computed with the trained MTP.
The difference between the LAMMPS and finite-difference results are insignificant, between 0 and 2 GPa.

\begin{table}
	\begin{center}
		\begin{tabular}{|c|c|c|c|} \cline{2-4}
			\multicolumn{1}{c|}{} & $C_{11}$ (GPa) & $C_{12}$ (GPa) & $C_{44}$ (GPa) \\ \hline
			MTP & 466 $\pm$ 22 & 161 $\pm$ 20 & 98 $\pm$ 14 \\ \hline
			DFT & 472\phantom{ $\pm$ 22} & 154\phantom{ $\pm$ 20} & 95\phantom{ $\pm$ 14}  \\ \hline
		\end{tabular}
		\caption{\label{tabl:ElasticConstants}
			Elastic constants of bcc-Mo as calculated by MTP and DFT.
			The 95\% confidence interval was given for the uncertainty of evaluating the elastic constants by MTP and well encloses the reference DFT results.}
	\end{center}
\end{table}

\section{Example 2: Melting point of Aluminum}\label{sec:example2}

Here we describe how to compute the melting point of aluminum by actively training an MTP on-the-fly.
The choice of Aluminum as a benchmark system is because of the availability of extremely accurate DFT results \cite{zhu2020-melting-point}.
The detailed description of the input, output, and intermediate files are given in the \verb|example-2/README| file of \cite{suppinfo-examples} and files referenced therein.

\subsection{Active learning iterations as implemented in MLIP}\label{sec:example2:active_learning}

We start with the description of the way the active learning iterations (Section \ref{sec:LOTF}) are implemented in MLIP, following Figure \ref{fig:lotf}.

\renewcommand{\step}[2]{\medskip\noindent{\bf Step #1 (#2).} }

\step{A}{active learning state preparation} The iteration starts by generating the \texttt{state.als} file based on the training set.
This is done by executing
\begin{verbatim}
mlp calc-grade pot.mtp train.cfg train.cfg temp.cfg
\end{verbatim}
This creates the \texttt{state.als} file containing the matrices described in Section \ref{sec:AL} which are needed for running LAMMPS with active learning. (The file \verb|temp.cfg| can be discarded.)

\step{B}{simulations} Now we use LAMMPS to run an MD simulation, switching on the active selection of configurations as indicated in the \verb|mlip.ini| file.
The extrapolative configurations are written to \texttt{preselected.cfg}.
According to the thresholds set in \verb|mlip.ini|.

\step{C}{selection} There may be a lot of configurations in \texttt{preselected.cfg}, especially when multiple MD trajectories are generated at Step B---in the later case their extrapolative configurations are merged into a single \texttt{preselected.cfg} file.
To select only non-repetitive, representative configurations, we use the \texttt{select-add} command:
\begin{verbatim}
mlp select-add pot.mtp train.cfg preselected.cfg selected.cfg
\end{verbatim}
This command forms the matrix \eqref{eq:matrix-B} and selects those configurations that maximize the determinant as explained in Section \ref{sec:AL}. The total number of selected configurations is guaranteed to be less than the number of parameters in \texttt{pot.mtp} (usually up to a few hundred) and is independent of the total number of configurations in \texttt{preselected.cfg}.

\step{D}{ab initio calculations} Now we calculate DFT energies, forces, and stresses for \texttt{selected.cfg} and write them to a new file, \texttt{computed.cfg}. In principle, this file may contain only a subset of configurations same as or close to the ones in \texttt{selected.cfg}---e.g., in the case if some DFT calculations were not done due to convergence issues or technical problems.
In this case missing configurations will be selected at the next iteration.\footnote{%
	The configurations at the next iterations will be different from the ones selected at the current iteration, due to the fact that the potential will be changed.
	In some applications this helps to solve issues arising from the lack of convergence of DFT self-consistent iterations, e.g., in cases when a configuration is at a cross-over between two magnetic states---if a configuration was at the cross-over, at the text iteration the selected configuration will be at a slightly changed.
}

\step{E}{merge} We append \texttt{computed.cfg} to \texttt{train.cfg} from the previous iteration. We consider this a separate step only to simplify the data flowchart in Figure \ref{fig:lotf}.

\step{F}{training} Finally, we re-fit the potential on the updated training set.
The produced potential, together with the expanded training set are the output of the iteration.
Considering all other files as intermediate ones, the net result of the iteration is the expanded training set and an updated potential that is able to make predictions in a larger configurational space.

\subsection{Stage 1: Active Learning of low-fidelity DFT during MD} \label{sec:stage_1}
Our first goal is to construct a potential trained on low-fidelity DFT calculations, whose purpose would be to be able to robustly sample configurations from solid and liquid phase of Al.
This is achieved in Stage 1.

Instead of starting the active learning cycle with an empty training set, we first generate an initial training set by running a 90-fs long VASP MD trajectory, adding every tenth configuration to the training set to avoid correlated configurations and save it to the file \texttt{train.cfg}.
We run an NVT-MD starting from an ideal 108-atom fcc configuration with the lattice parameter of 4.1 \AA.
Only $\Gamma$-point was used for the k-point integration and the energy cutoff of 410 eV was used for the plane-wave basis.
The training set, thus, contains 10 configurations.

We choose the potential of level 16 with cutoff of $5\,\angstrom$, \verb|mindist| of $2\,\angstrom$, and the radial basis size of 6, and save it to \texttt{init.mtp}.
We next train this potential on the database \texttt{train.cfg} with the following command:
\begin{verbatim}
mlp train init.mtp train.cfg --trained-pot-name=pot.mtp
\end{verbatim}
The potential \texttt{pot.mtp} and training set \texttt{train.cfg} are the input to the active learning iteration.

\subsubsection*{Stage 1a: active learning on a single MD trajectory}

We now run active learning iterations, in which we run molecular dynamics and collect configurations on which the potential attempts to extrapolate, in accordance to the general workflow shown in Figure \ref{fig:lotf} and detailed in Section \ref{sec:example2:active_learning} focusing on data processing aspect of an active learning iteration.
We set the thresholds $\gamma_{\rm select} = 3$ and $\gamma_{\rm break} = 10$.
We now describe the steps of the active learning iteration.

We actively train an MTP while running an NVT-MD under the same setting as the initial AIMD trajectory---at 900 K starting from a 108-atom fcc configuration with the lattice parameter of 4.1 \AA.
The initial velocities were kept the same in order to ensure that the MD trajectory on subsequent iterations are close to those on the previous iteration---which in turn was done solely for illustration purposes to see the gradual increase in the number of time steps an MD with MTP can reliably do.

During the first iteration the MTP successfully recognizes that the initial configuration is in the training set, makes a time step and selects the next configuration to be added to the training set.
(LAMMPS and VASP initialize the MD with different velocities, hence the trajectories and different.)
On the second iteration, four time steps were made until the simulation exceeded the upper threshold, and two configurations (sampled from the last two time steps) were added to the training set.
This process continued until the 58th iteration, on which a 10-ps MD was run and no configuration was selected as extrapolative.
Totally, 225 configurations were actively selected during the entire process by the 58th iteration.

\begin{figure}[ht]
	\centering
	\includegraphics{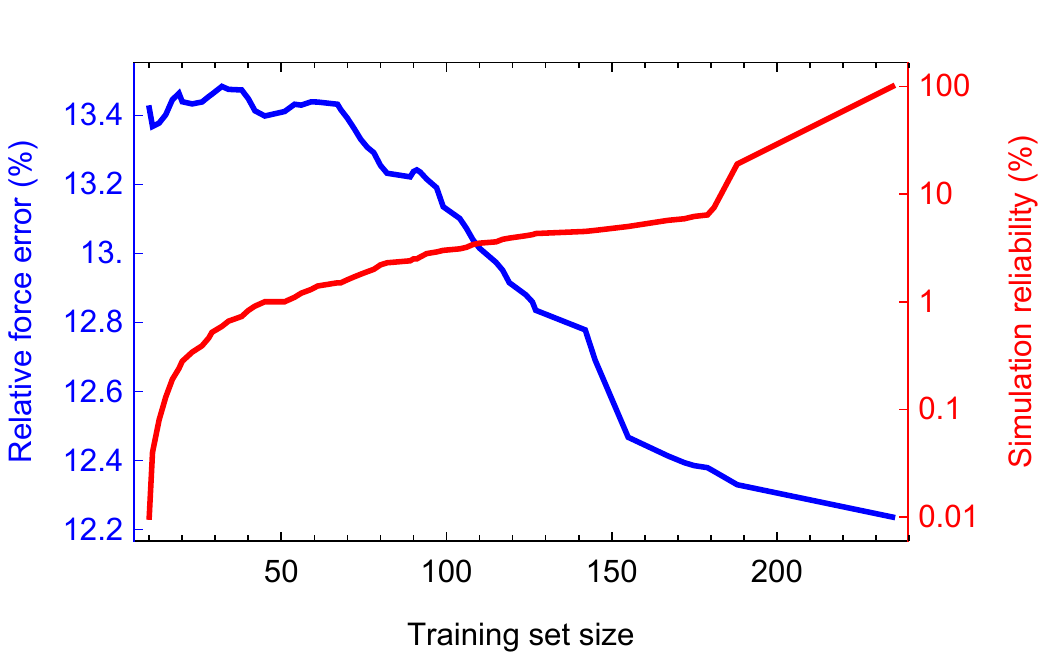}
	\caption{%
		Performance of active learning bootstrapping iterations on Stage 1a.
		The reliability is defined as the number of time steps made before exceeding $\gamma_{\rm break}$ relative to the total number of time steps (10\,000).
		One can see that the error drops only slightly---from 13.5\% to 12\%.
		As will be shown, the large error is mostly due to the k-point noise in the data that will be removed on the text stage.
		The major effect of the iterations is reliability---it increases from 0 to 100\% after reaching 235 configurations in the training set.
	}
	\label{fig:example-stage1a}
\end{figure}

To better understand the performance of the active learning iterations, we generated a validation set of 200 configurations sampled at 900 K and computed them with DFT.
We emphasize that this validation set is typically not needed in practice in the active learning iterations other than for testing purposes.
The validation error and the number of time steps of MD at each iteration is plotted in Figure \ref{fig:example-stage1a}.
The number of time steps is a good indicator of the reliability of the potential.
One can see that even a potential trained on the first 10 configurations has a good validation error of 13.5\% which in the course of active training goes down to only 12\%.
The 10 configurations have altogether 1080 forces which appears to be sufficient for reaching a good accuracy on a potential with about 120 parameters.
However, the algorithm requires extra 225 configurations to attain the perfect reliability of the simulation.

The resulting 235 configurations are rather correlated---they were collected from essentially the same MD trajectory approximately every 10 time steps.
Hence on the 59th iteration we sparcify the training set: we select only those configurations with respect to which other configurations are interpolative.
We do it with the \verb|select-add| command.
It selects 78 configuration that we keep for the subsequent iterations.

\subsubsection*{Stage 1b: active learning during multi-scenario MD simulations}

Next we perform the 60th iteration in the same manner as iterations 1--58 except that we run 60 LAMMPS trajectories in parallel, with the lattice parameter $a$ between 4.0 and 4.25 \AA\ and temperatures $T$ between 700 and 1100 K for two scenarios: \texttt{fcc} and \texttt{liquid}.
The \texttt{fcc} scenario is exactly the same as described above, and in the liquid scenario we first run 10\,000 steps with $a=4.25\,\angstrom$ and $T=1500\,K$, then we run 10\,000 steps with the target temperature (between 700 and 1100 K) gradually reducing $a$ to the target value, and finally run 10\,000 steps with the target temperature and lattice parameter.
The \texttt{preselected.cfg} files from these trajectories are concatenated into one file before doing the selection step.
The thresholds are set $\gamma_{\rm select}=\gamma_{\rm break}=5$.
Equal thresholds ensure that only one configuration will be selected from each trajectory---which will ensure that not too many configurations will be given to the \texttt{select-add} command.

Again, we have generated a validation set of 180 configurations corresponding to the above 60 MD scenarios, in order to understand how the active learning iterations perform.
Before the 60th iteration the measured force error was 20\%.

\begin{figure}[ht]
	\centering
	\includegraphics{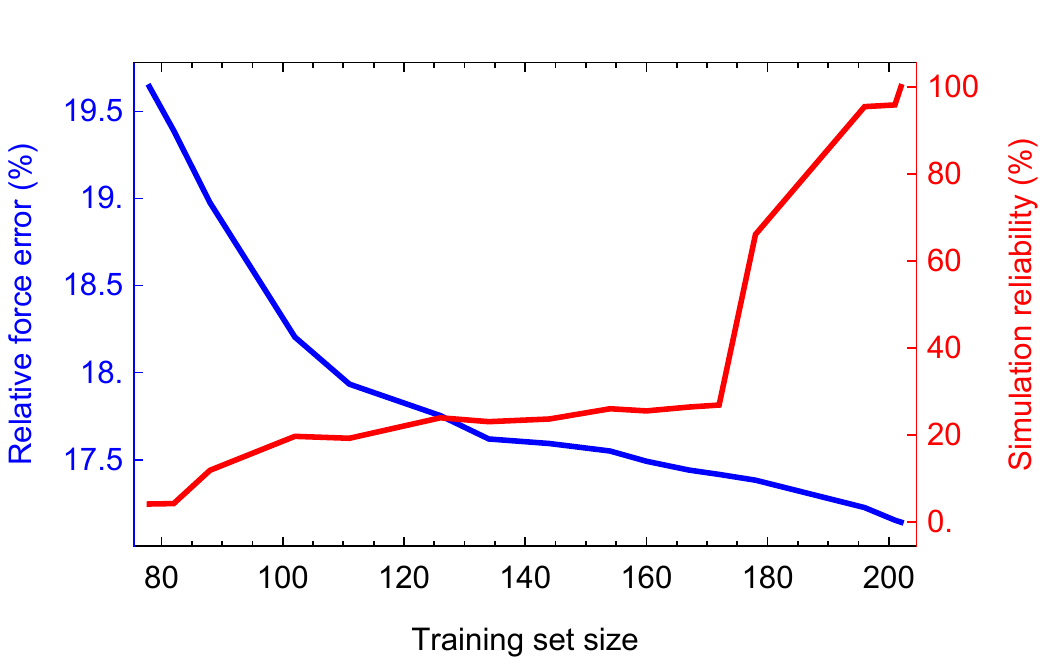}
	\caption{%
		Performance of active learning bootstrapping iterations on Stage 1b.
		The reliability is defined as the number of time steps made before exceeding $\gamma_{\rm break}$ relative to the total number of time steps.
		As in Stage 1a, the large error is mostly due to the k-point noise in the data that will be removed on the text stage.
		Likewise, the major effect of the iterations is reliability---it increases from 5 to 100\% after reaching 202 configurations in the training set.
	}
	\label{fig:example-stage1b}
\end{figure}

This was repeated for 14 additional iterations, until at the 74th iteration no configurations were selected for training.
The size of the training set reached 202 configurations by the 75th iteration, which means that total 359 static DFT calculations were made.
We note that because only the $\Gamma$-point was used, these 359 were rather cheap and did not take the majority of computational time during this process.
The force error decreased from 20\% to 17\%.
The performance over iterations 60--74 is plotted on Figure \ref{fig:example-stage1b}.

Finally, on the 75th iteration we use the trained potential to sample 186 solid and liquid configurations that we later compute with high-accuracy DFT.
To that end, we switch off active learning (option \texttt{select FALSE} in the \texttt{mlip.ini} file) in the LAMMPS simulations.
This set was generated similarly to the validation set, but with extra six configurations corresponding to liquid at $T=1500$ K and $a=4.25$---we use these parameters to prepare the liquid state in all liquid simulations.

\subsection*{Stage 2: High-Accuracy DFT}

This stage starts with 186 liquid and solid configurations computed with the $3\times 3\times 3$ k-points mesh.
We fit five MTPs to this training set, with random initialization of parameters, and select the one with the lowest error---this is the beginning of the new round of active learning iterations.
We use the same 180-configuration validation set computed with the new DFT parameters to assess the error.
The error immediately drops from 17\% to 4.5\%---this is the result of improving the accuracy of the data.
This means that on Stage 1 the magnitude of the error was high mostly due to the k-point noise in data.

We next run the same iterations as in Stage 1b: the only difference is a more accurate k-point mesh in the DFT calculations on Step D.
We repeat the active learning simulations until we obtain a potential with which we run a molecular dynamics and select no extrapolative configurations. We use this potential for the last stage.

\begin{figure}[ht]
	\centering
	\includegraphics{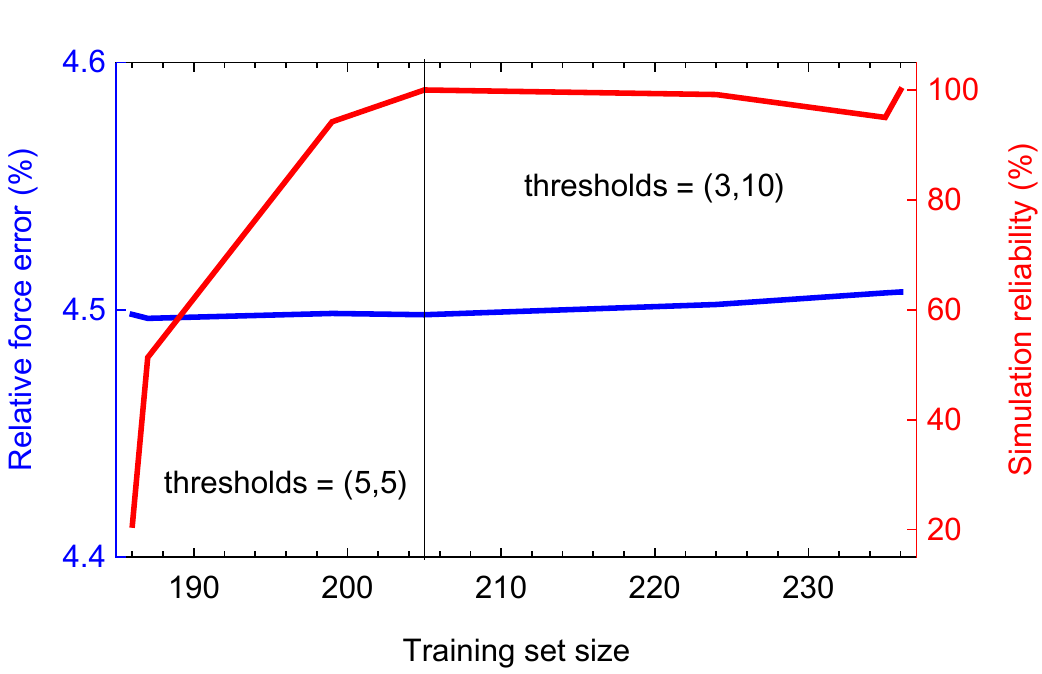}
	\caption{%
		Performance of active learning bootstrapping iterations on Stage 2.
		The reliability is defined as the number of time steps made before exceeding $\gamma_{\rm break}$ relative to the total number of time steps.
		The slight decline in reliability is because of how it is measured---$\gamma_{\rm break}$ was reduced to a tighter value after reaching 205 configurations in the training set.
		The error almost does not change.
		As before, the major effect of the iterations is reliability.
	}
	\label{fig:example-stage2}
\end{figure}

\subsection*{Stage 3: Coexistence simulations}

At this stage we completely switch off active learning and perform the standard coexistence simulations with LAMMPS.
We start by running four molecular dynamics simulation of solid and liquid with $4\times12^3 \approx 7000$ atoms in a supercell with $T\in\{900\, K,1000\, K\}$.
Without active learning, LAMMPS with MLIP can be run in the MPI-parallel mode, efficiently accelerating the simulation of 7000 atoms.
We use them to fit the fcc lattice parameter as a function of temperature, and the coefficient of expansion of the ``long axis'' for the 50\%--50\% solid-liquid coexistence system.

We then set up a coexistence simulations in the following way: (0) we start by a $12\times12\times48$ system as measured in the lattice parameter units. We run 3000 steps of the NVT with the target temperature $T_{\rm target}$ and the corresponding lattice parameter, then we freeze the lower half ($12\times12\times24$) of the system and run an MD simulation at 1200 K, expanding the supercell in the third axis until the supercell volume will correspond to the target temperature.
Finally, we reset the velocities of all the atoms to a lower temperature $T_{\rm reset}$, to compensate for the higher energy of the liquid, and run a classical MD with the NVE ensemble.
After a few runs we found that $T_{\rm target} = 885$ K and $T_{\rm reset} = 610$ K yields a 50\%--50\% coexistence, with the axial stresses less than 0.02 GPa by absolute value and coexistence temperature of 885 K.
In other words, the computed melting temperature of Al on the 3x3x3 k-point mesh is 885 K, just 3 K smaller than the extremely accurate DFT calculation of \cite{zhu2020-melting-point}.

\section{Example 3: Stable convex hull of Ag-Pd structures} \label{sec:example-3}

In the last example we show how to calculate the convex hull of the Ag-Pd binary system using active learning. Constructing a convex hull means identifying the most stable (with the lowest formation energy) binary structures with the composition Ag$_{1-x}$Pd$_x$ with $x$ between 0 (pure Ag) and 1 (pure Pd). 
We identify the stable structures by selecting among the finite number of the so-called ``candidate structures'' those ones that lie on the lower convex hull on the energy-composition plot.
As for the previous examples, the files mentioned below are available at \cite{suppinfo-examples}.

The structures themselves may be of different origin, but usually they are provided by some generative algorithm (e.g., \cite{glass2006uspex}) or are taken from some bank of structures (e.g, \cite{ong2015materials}, \cite{curtarolo2012aflow}). We follow the second way: generate 39k crystal structures with up to 12 atoms and different underlying lattice types (fcc, bcc, hcp) populated with different numbers of Ag and Pd atoms to introduce different concentrations. 

The samples then undergo structure relaxation (i.e., energy minimization) to relieve interatomic forces and lattice stresses by changing (relaxing) lattice vectors and atomic positions. Similarly to the MD simulation from Section \ref{sec:example2:active_learning}, each relaxation produces a trajectory starting with the structure to be relaxed and ending with the ``relaxed'' one having practically zero forces and stresses.
The active learning was used in this scenario as well, with the source of new configurations being the relaxation trajectories of candidate structures. 

To construct a convex hull for the Ag-Pd system (simultaneously fitting a corresponding moment tensor potential) we launch the active learning procedure as described in Section \ref{sec:example2:active_learning}, this time starting with an empty training set.
We start by choosing the same potential as in Section \ref{sec:stage_1}.

The only essential difference of the active learning workflow compared to Section \ref{sec:example2:active_learning} is that the step B (simulation) is replaced by relaxation with the MLIP package.
During this step we use MLIP to run relaxation for each candidate structure. Each relaxation can end up in two ways: it can either finish successfully producing an equilibrium configuration; or terminate early if an extrapolative configuration occurs in the relaxation trajectory. 
The successfully relaxed structures are appended to the file \texttt{relaxed.cfg}.
Active learning iterations keep going till all the relaxation have finished successfully providing the corresponding equilibrium structures---in our case this happened on the 5th iteration.

\begin{figure}[ht]
	\begin{subfigure}{.55\textwidth}
		\centering
		\includegraphics[width=.8\linewidth]{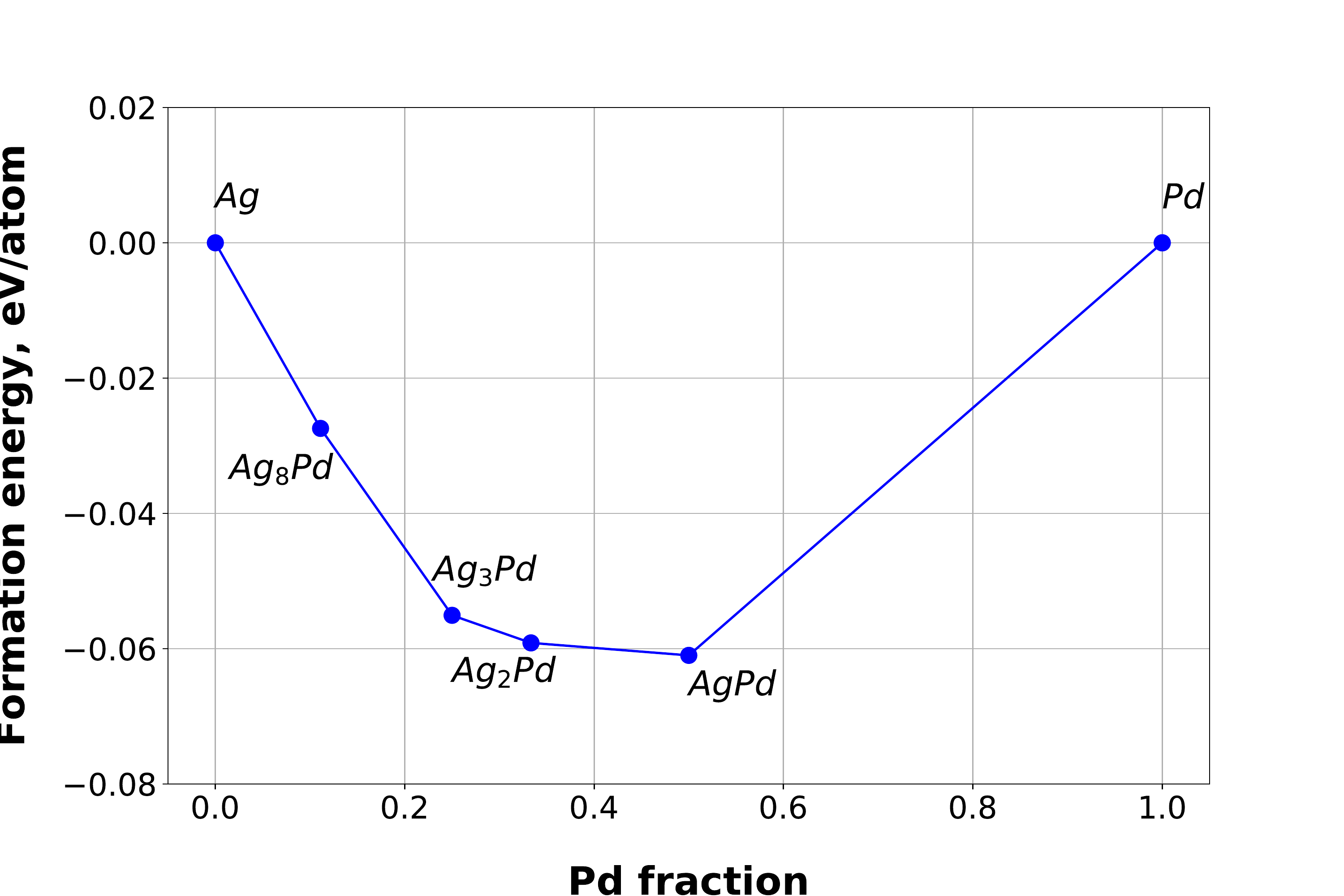}  
		\caption{Convex by DFT}
		\label{fig:DFT_CH}
	\end{subfigure}
	\begin{subfigure}{.55\textwidth}
		\centering
		\includegraphics[width=.8\linewidth]{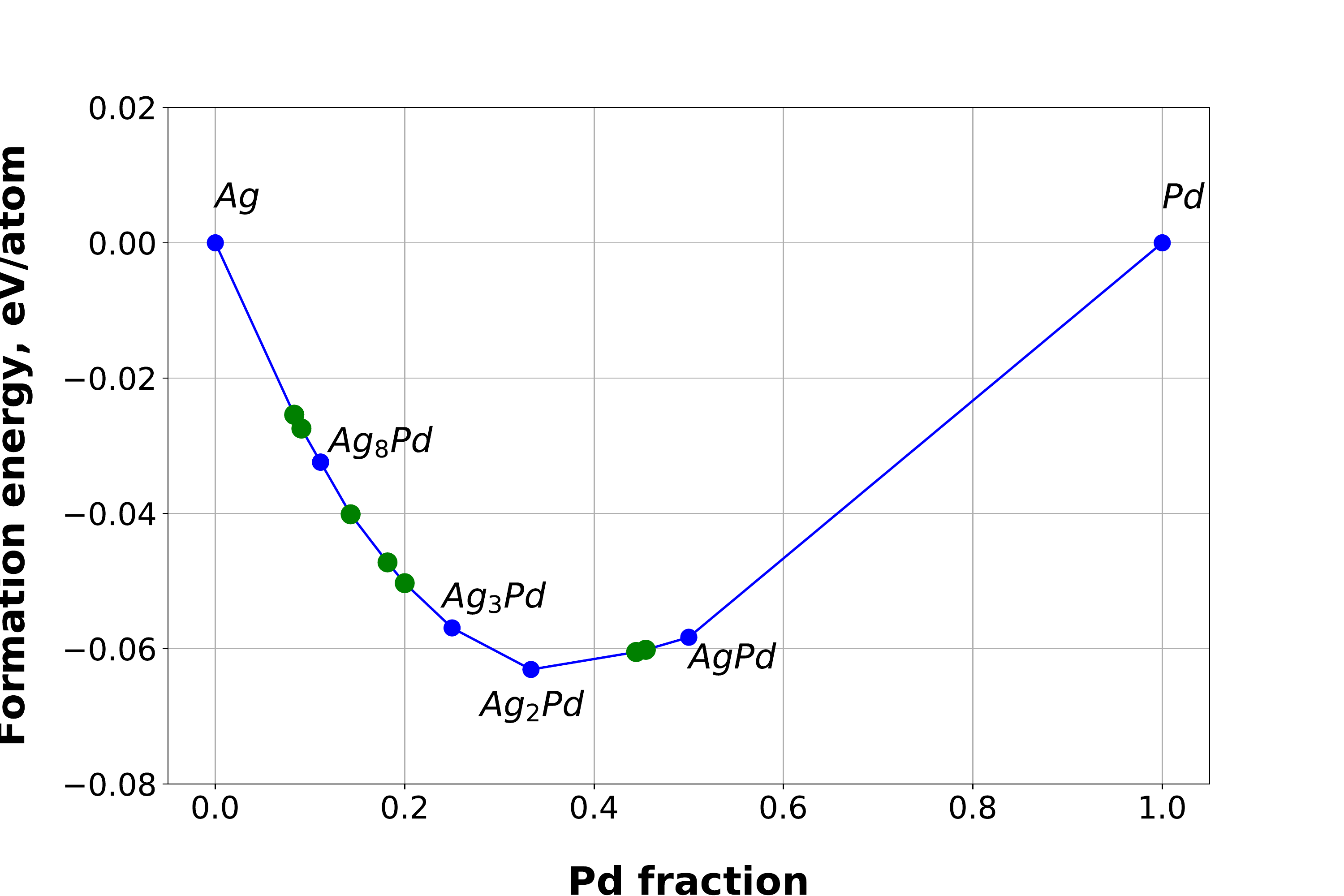}  
		\caption{Convex by MTP}
		\label{fig:MTP_CH}
	\end{subfigure}
	\caption{Convex hull obtained by MTP includes more structures (green dots) compared to the one obtained by DFT, thanks to 100 times more candidate structures used.}
	\label{fig:CH_vasp_vs_mtp}
\end{figure}

After the iterations end, the last iteration produces the \texttt{relaxed.cfg} file containing all the configurations successfully relaxed. Based on the energies of the relaxed structures provided by MTP, their formation energies are calculated, which allows then to construct a convex hull (see Fig. \ref{fig:MTP_CH}).

When the last active learning iteration has ended, the training set included 442 configurations on which MTP has the MAE and RMSE of 1.9 meV/atom and 2.4 meV/atom, respectively.
The effect of this error may be that the ground-state structures happen to slightly miss the MTP-based convex hull as they may rise above other structures in energy.

It is hence interesting to check the MTP results by constructing a convex hull based on 302 candidate structures from the Aflow library \cite{curtarolo2012aflow}, and relaxing them on DFT (as implemented in VASP 5.4.4).
We relax those structures to make sure the same DFT settings were used to train MTP and to compute the DFT convex hull.
The convex hull constructed from these structures is shown on Fig. \ref{fig:DFT_CH}. 
The results of this comparison are in favor of MTP: it includes all the DFT ground-state structures, despite the 2 meV/atom RMS error in the energy.
If one needs to completely eliminate the MTP errors from the resulting convex hull, one may post-relax some of the most stable structures on DFT (as was done in \cite{gubaev2019-alloys}).
Some extra structures, not included in the DFT convex hull, appear in the MTP convex hull due to a better coverage of the compositional space by the 39k initial structures used by MTP.

In this example the MTP shows its predictive power, as it has almost precisely reproduced the results obtained through DFT relaxation of very carefully chosen samples, but has fulfilled this task by relaxing a general system-agnostic pool of candidate structures.
The benefits from using an active learning approach consist of a several-fold savings in DFT calculations---442 static calculations are much cheaper than 302 relaxations especially considering that these calculations are typically restarted several times to avoid the error of originating from the plane-wave basis being fixed during the DFT relaxation.
The computational cost of training and relaxing of the 39k samples is less than 10\% and the error of 2meV/atom is comparable to discrepancies in formation energies caused by different DFT pseudopotentials, or even different convergence parameters.

\section{Concluding Remarks}

This manuscript gives a formulation of the moment tensor potentials (MTPs) and the active learning algorithm, outlines the structure of the MLIP code implementing MTPs and active learning, and provides three detailed examples of the usage of MLIP to perform particular atomistic simulations.
Our goal was that the simulations could be easily reproduced, therefore the examples were chosen sufficiently familiar for practitioners, yet advanced enough to contain many of the components of simulations that the frontiers of computational materials science need.
The supplemental information \cite{suppinfo-examples} contain all the files and data needed to reproduce the results presented in the manuscript.

\section{Acknowledgements}

This work was supported by the Russian Science Foundation (grant number 18-13-00479).

\bibliographystyle{plain}
\bibliography{paper}

\end{document}